# Anomalous Raman Scattering from Phonons and Electrons of Superconducting FeSe$_{0.82}$

Pradeep Kumar[1], Anil Kumar[2], Surajit Saha[1], D. V. S. Muthu[1], J. Prakash[3], S. Patnaik[4], U. V. Waghmare[2], A. K. Ganguli[3], and A. K. Sood[1,*]

[1]Department of Physics, Indian Institute of Science, Bangalore -560012, India

[2]Theoretical Sciences Unit, Jawaharlal Nehru Centre for Advanced Scientific Research, Bangalore -560064, India

[3]Department of Chemistry, Indian Institute of Technology, New Delhi -110016, India

[4]School of Physical Sciences, Jawaharlal Nehru University, New Delhi - 110067, India

## Abstract

We report interesting anomalies in the temperature dependent Raman spectra of FeSe$_{0.82}$ measured from 3K to 300K in the spectral range from 60 to 1800 cm$^{-1}$ and determine their origin using complementary first-principles density functional calculations. A phonon mode near 100 cm$^{-1}$ exhibits a sharp increase by ~ 5% in frequency below a temperature $T_s$ (~ 100K) attributed to strong spin-phonon coupling and onset of short-range antiferromagnetic order. In addition, two high frequency modes are observed at 1350 cm$^{-1}$ and 1600 cm$^{-1}$, attributed to electronic Raman scattering from ($x^2$-$y^2$) to $xz$ / $yz$ $d$-orbitals of Fe.

## 1. INTRODUCTION

Recent discovery of superconductivity in rare earth iron pnictides $RFeAsO_{1-x}F_x$ (R = La, Sm, Ce, Nd, Pr, and Gd [1-3]) has now expanded to include the related alkali doped $A_xM_{1-x}Fe_2As_2$ (A = K, Na; M = Ca, Sr, Ba) [4-5] and iron-chalcogenides $Fe_{1+\delta}Se_{1-x}Te_x$ [6-7]. Iron pnictides and iron chalcogenides share the common feature of tetragonal *P4/nmm* structure with FeAs or FeSe layers, wherein Fe is tetrahedrally coordinated with As or Se neighbours. The superconducting transition temperature ($T_c$) for iron chalcogenides has increased from initial 8K [6] to 14K [7] with suitable amount of Te substitution, and to 27K [8] under high pressure. An interesting observation in the $Fe_{1+\delta}Se_{1-x}Te_x$ system is the occurrence of a structural phase transition from tetragonal to orthorhombic at $T_s \sim 100K$ [6, 9], accompanied by an anomaly in magnetic susceptibility [7]. Wang et al. [10] have shown that only the thin films of tetragonal $FeSe_{1-x}$, which show such a low temperature structural transition are superconducting, thus suggesting a crucial link between structural transition and superconductivity. Earlier reports of superconductivity in tetragonal $FeSe_{1-x}$ compound termed PbO- tetragonal phase as α-phase [6]. However, some other reports have termed the PbO-tetragonal phase as β-FeSe as well [11, 12], causing some confusion in the nomenclature. A recent report of the phase diagram [13] (where the tetragonal phase is designated as β-phase), claims that only compounds with the stoichiometry in the narrow range of $Fe_{1.01}Se$ to $Fe_{1.03}Se$ are superconducting. Understanding the primary origin of these co-occurring structural transition and magnetic anomaly at $T_s$ is essential in uncovering important couplings in the normal phase that are relevant to superconductivity in $FeSe_{1-x}$ and related materials.

In this paper, we report Raman scattering from tetragonal FeSe$_{0.82}$ with onset $T_c$ of ~ 12K. To the best of our knowledge, only room temperature Raman spectra associated with phonons has been reported for tetragonal FeTe and Fe$_{1+\delta}$Se$_{1-x}$Te$_x$ [14]. There are two motivating factors behind this work: first to identify the phonons relevant to the observed structural phase transition, and use first-principles calculations to determine the possible coupling between these phonons and accompanying changes in magnetic structure via spin-phonon coupling, if any. This may throw light on the possible role of electron-phonon coupling through spin channel in the mechanism of superconductivity [15]. Secondly, to determine the changes in electronic states and excitations near Fermi level through the transition(s) at $T_s$, which may set the stage for superconductivity at lower temperatures. As emphasized earlier [16], Fe orbitals contribute significantly to the electronic states near the Fermi level and hence they are expected to play a crucial role in the mechanism of superconductivity. In particular, it is not clear experimentally if $d_{xz}$ and $d_{yz}$ orbitals are split or not [17-20].

Here, from the temperature dependence of Raman scattering from phonons and electrons (in crystal-field split $d$-orbitals of Fe) in tetragonal-FeSe$_{0.82}$, we present two significant results: (i) The lowest frequency phonon, associated with Se vibration in the *ab* plane, shows anomalously large blue shift of ~ 5% in frequency below $T_s$; (ii) The $d_{xz}$ and $d_{yz}$ orbitals of Fe are non-degenerate with a splitting of ~ 30 meV. This is the first experimental evidence suggesting the possible splitting of $d_{xz}$ and $d_{yz}$ energy levels that is being discussed in recent theoretical studies [17-20] in the general class of iron based superconductors. In ref [19], related to LaFeAsO, the splitting of 3d orbitals of Fe results from combined effects of tetrahedral crystal field, spin-orbit coupling, strong hybridization between Fe $d$ and As $4p$ orbitals and lattice compression along z-axis. Also, it has been suggested [20] that, since $d_{xz}$ and $d_{yz}$ orbitals are roughly half filled, spontaneous symmetry breaking can lead to the lifting of the degeneracy of $d_{xz}$ and $d_{yz}$ orbitals. Using first-principles Density Functional Theory (DFT) based calculations; we show that our experimenatl observations can be readily understood within a single picture: there is a transition to a phase with short-range stripe [21] antiferromagnetic order below $T_s$, with orthorhombic structural distortion as a secondary order parameter.

To put our studies in perspectives, we note that there are a few temperature dependence Raman studies on $NdFeAsO_{1-x}F_x$ [22, 23], $Sr_{1-x}K_xFe_2As_2$ (x = 0 and 0.4) [24], $CaFe_2As_2$ [25], $R_{1-x}K_xFe_2As_2$ ( R = Ba, Sr) [26, 27] and $Ba(Fe_{1-x}Co_x)_2As_2$ [28]. In case of $NdFeAsO_{1-x}F_x$ [22-23] none of the observed phonon modes show any anomaly below the superconducting transition temperature. Similarly no anomaly was seen as a function of temperature in $Sr_{1-x}K_xFe_2As_2$ [24]. However, in another study of $R_{1-x}K_xFe_2As_2$ ( R= Ba, Sr) [26] the linewidths of the phonon modes involving Fe and As near 185 cm$^{-1}$ ( $A_{1g}$ ) and 210 cm$^{-1}$ ( $B_{1g}$ ) show a significant decrease below the spin-density-wave transition temperature $T_s$ ~ 150K, arising from the opening of the spin-density-wave gap. Also, the phonon frequency of the 185 cm$^{-1}$ mode shows a discontinuous change at $T_s$, signaling the first order structural transition accompanying the spin-density-wave transition at $T_s$. Similar results for the $B_{1g}$ mode ( near 210 cm$^{-1}$) are seen for $Sr_{0.85}K_{0.15}Fe_2As_2$ and $Ba_{0.72}K_{0.28}Fe_2As_2$ ($T_s$ ~ 140K) [27]. In parent compound $CaFe_2As_2$, the $B_{1g}$ phonon frequency ( 210 cm$^{-1}$) shows a discontinuous decreases at $T_s$ ~ 173K and $A_{1g}$ phonon ( near 190 cm$^{-1}$ ) intensity is zero above $T_s$ , attributed to the first order structural phase transition and a drastic change of charge distribution within the FeAs plane[25]. The $E_g$ (Fe, As) phonon (~ 135 cm$^{-1}$ ) in Ba(Fe$_{1-x}$Co$_x$)$_2$As$_2$ ( x < 0.6) splits in two modes near the structural transition temperature ( $T_s$ ~ 100 to 130K) linked to strong spin-phonon coupling [28].

## 2. METHODS

### 2.1. Experimental Details

Polycrystalline samples with nominal composition of FeSe$_{0.82}$ were synthesized by sealed tube method (690 C/24h and 720 C/24h) using the elements as starting materials. The powder X-ray diffraction pattern of the compound shows the presence of the tetragonal FeSe phase (space group *P4/nmm*) with a very small amount of hexagonal FeSe (NiAs type) phase (< 5%) (Diffraction lines marked by * in Fig. 1). The onset of superconducting transition occurs at 11.8 K (see inset b of Fig.1).

The resistivity decreases with temperature and the onset of superconducting transition occurs at 11.8 K with residual resistivity value (RRR = $\rho_{300}/\rho_{13}$) of 4.9. The criterion used for the determination of $T_c$ is shown in the inset (b) of Fig. 1. The inductive part of the rf magnetic susceptibility attesting the onset of bulk diamagnetic state with transition temperature near 11.8 K is shown in the inset (c) of Fig. 1.

Unpolarised micro-Raman measurements were performed on sintered pellets of FeSe$_{0.82}$ in backscattering geometry, using 514.5 nm line of an Ar-ion Laser (Coherent Innova 300), at 20 different temperatures from 3K to 300K using a continuous flow liquid helium cryostat with a temperature accuracy of ± 0.1K. The scattered light was analysed using a Raman spectrometer (DILOR XY) coupled to liquid nitrogen cooled CCD detector.

### 2.2 First-principles Calculations

Our first-principles calculations are based on DFT as implemented in PWSCF [29] package. We use ultrasoft pseudopotentials [30] to describe the interaction between the ionic cores and the valence electrons, and a plane wave basis with energy cutoffs of 40 Ry for wave functions and 480 Ry for charge density. To model Se vacancies, we use $\sqrt{2} \times \sqrt{2} \times 1$ and $2 \times 2 \times 1$ super cells of $FeSe_{1-x}$ for x = 0.00 and 0.0125 in $FeSe_{1-x}$, respectively. Structural optimization is carried through minimization of energy using Hellman-Feynman forces and the Broyden-Flecher-Goldfarb-Shanno based method. Frequencies of the zone centre (q = 0,0,0) phonons are determined using a frozen phonon method for the relaxed structure with experimental lattice constants. We have also carried out calculations to include effects of correlations, likely to be important in these iron pnictides [31], with an LDA+U correction, and find that phonon frequencies do change sizably with U [32], confirming the conclusion of ref. [31] in the context of phonons. However, the picture (relevant to Raman measure here) developed here based on LDA calculations does not change qualitatively.

## 3. RESULTS AND DISCUSSION

### 3.1 Raman Scattering

There are four Raman active phonon modes belonging to the irreducible representations $A_{1g} + B_{1g} + 2E_g$ [14]. Figure 2 shows Raman spectrum recorded at 3K where the spectral range is divided into two parts: low frequency range (80 to 350 $cm^{-1}$) showing 5 Raman bands, labeled as S1 to S5 and high frequency range (1200 to 1800 $cm^{-1}$) displaying weak Raman bands, S6 and S7. Following the recent Raman studies of tetragonal-FeTe and $Fe_{1.03}Se_{0.3}Te_{0.7}$ [14] and our calculations (to be discussed below), the four Raman active mode are S1: 106 $cm^{-1}$, symmetry $E_g$, mainly Se vibrations; S2: 160 $cm^{-1}$, $A_{1g}$, Se vibrations; S3: 224 $cm^{-1}$, $E_g$, Fe vibrations, and S4: 234 $cm^{-1}$, $B_{1g}$, Fe vibrations. Mode S5 is weak and its intensity depends on the spot on the sample. We assign the mode S5 (254 cm$^{-1}$) to the hexagonal phase of FeSe [33].

Raman spectra were fitted with a sum of Lorentzian functions to derive the mode frequencies and linewidths. The linewidths of all the modes show normal temperature dependence i.e., increase by ~ 3 cm$^{-1}$ from 3K to 300K and are not shown. It can be seen in Fig. 3 that while the frequencies of modes S2, S3 and S4 show normal temperature dependence, mode S1 shows a sharp change near $T_s$ ~ 100K, where a tetragonal to orthorhombic phase transition is expected [6, 7, 9]. The frequency $\omega(T)$ of the S1 mode increases by a large amount by ~ 5% below $T_s$. The solid lines in Fig. 3 are fitted to an expression based on cubic anharmonicity where the phonon mode decays into two equal energy phonons [34]: $\omega(T) = \omega(0) + C\{1+2n[\omega(0)/2]\}$, C is Self-energy parameter for a given phonon mode and $n(\omega) = 1/[\exp(\hbar\omega/k_BT) -1]$ is the Bose-Einstein mean occupation factor. The fitting parameters $\omega(0)$ and C are given in Table I. We will come back to discuss the anomalous change in frequency of the S1 mode after presenting density functional calculations. The lineshape of the S1 mode is Lorentzian and not Fano as shown in the inset (a) of Fig. 2 implying negligible or no coupling of this phonon with electrons.

In addition to the expected Raman active phonon modes, Fig.2 shows two additional modes S6 (1350 cm$^{-1}$) and S7 (1600 cm$^{-1}$). These modes are very weak at all temperatures and hence their quantitative temperature dependence was not studied. In earlier Raman studies [35] of CeFeAsO$_{1-x}$F$_x$ at room temperature, weak Raman modes have been noticed at 846 and 1300 cm$^{-1}$. It was suggested [35] that the high frequency modes may be related to the electronic Raman scattering involving the *d*-orbitals of Fe.

We attempt to understand S6 and S7 modes in terms of electronic Raman scattering involving Fe *d*-orbitals. In the case of x = 0.0, Fe atoms are tetrahedrally coordinated with Se atoms and the crystal field (CF) leads to the splitting of *d*-orbitals of Fe into four energy levels $x^2-y^2$, $z^2$, *xy* and *xz/yz* (*xz* and *yz* orbitals remain degenerate). Figure 4(a) shows the energy diagram as given in ref [17, 18]. The mode at 1350 cm$^{-1}$ is very close to the energy difference between $x^2-y^2$ and *xz/yz*. Our clear observation of two modes (S6 and S7) with an energy difference of δE ~ 240 cm$^{-1}$ (30 meV) suggests that the *d*-orbitals *xz* and *yz* are split with this energy difference as theoretically argued in ref [19-20]. This is schematically shown in Fig. 4(b).

**3.2 First-Principles Simulations**

For the non-magnetic (NM) state of FeSe, our optimized internal structure agrees within 0.5 % with the experimental one, and estimates of phonon frequencies are slightly lower for x = 0.125 as compared to those of undoped FeSe (see Table I). While the discrepancy of the calculated and measured values of phonon frequencies is larger than the typical errors of DFT calculations, it is comparable (or slightly better) to that in a recent DFT calculation of phonons in tetragonal-FeTe [14]. This could originate from various factors such as change in magnetic ordering, electronic correlations, disorder in Se vacancies and possibly strong spin-phonon couplings in the FeSe$_{1-x}$.

Interestingly, our calculations on FeSe with initial state of ferromagnetic (FM) or antiferromagnetic (opposite spins at nearest neighbors, AFM1) ordering of spins at Fe sites converged to a NM state upon achieving self-consistency. On the other hand, antiferromagnetically ordered stripe [21] phase (AFM2) of FeSe is stable and lower in energy than the NM phase by 50 meV/cell. In the stripe AFM2 phase, $S_i \cdot S_j$ vanishes for the nearest neighbors (NN) and is negative for the second nearest neighbor (SNN) spins. The stability of AFM2 and instability of FM and AFM1 states can be understood if the exchange coupling between SNN spins is antiferromagnetic and that between NN spins is ferromagnetic of roughly the same magnitude: FM and AFM1 states are frustrated. The origin of this can be understood through analysis of super-exchange interaction between Fe sites, mediated via p-states of Se. Opposite signs of the exchange coupling for NN and SNN spins arise from the fact that Fe-Se-Fe angles relevant to their super-exchange interactions are closer to 90 and 180 degrees, respectively. Our DFT calculations for $x \neq 0$ reveal that the AFM1 state is locally stable as the magnetic frustration is partially relieved by Se vacancies. From the energy difference between AFM1 and AFM2 states (for $x \neq 0$), we estimate the exchange coupling to be about 10 meV. Due to frustrated magnetic interactions, we expect there to be many states close in energy to the stripe AFM2 state. To this end, we carried out Monte Carlo simulations with a 2-D Ising model upto second nearest neighbor interactions ($-J_1 = J_2 = 10$ meV), and find that the system develops only *short-range order* of the stripe kind [32] below about 100 K and that statistical averages are rather demanding numerically due to frustrated J-couplings.

Interestingly, our calculation shows that the tetragonal symmetry of FeSe is broken with this stripe-AFM order, as manifested in the stresses $\sigma_{xx} \neq \sigma_{yy}$ for the AFM-stripe phase, whereas $\sigma_{xx} = \sigma_{yy}$ for the NM phase. A weakly orthorhombic symmetry of the stress on the unit cell induced by the AFM stripe order on the unit cell is expected to result in orthorhombic strain, $\frac{b-a}{a} \neq 0$, in the crystal as a secondary order parameter in a transition to AFM-stripe phase, similar to that in improper ferroelectrics (we note that there are no unstable phonons in the nonmagnetic tetragonal phase ruling out a primarily structural transition).

We now attempt to understand the large observed change of ~ 5% in the frequency of the S1 ($E_g$) mode below $T_s$. To this end, we first determined the change in frequency of phonons as a function of orthorhombic strain within DFT. It has been shown experimentally that maximum value of (b-a)/a is ~ 0.5% at the lowest temperature of 5K [9]. Our calculations show that phonon frequencies hardly change with an orthorhombic strain of ~ 0.5% in the non-magnetic state, indicating a weak strain-phonon coupling and ruling out its role in T-dependence of the S1 mode frequency. Thus, it has to be a strong spin-phonon coupling that is responsible for the anomalous behavior of the mode S1 below $T_s$. To this effect, our DFT calculations show that the mode S1 frequency in $FeSe_{0.875}$ indeed hardens by ~ 7 % (to 126 cm$^{-1}$) in the AFM stripe phase as compared to that (118 cm$^{-1}$) in the AFM1 state. This hardening is a theoretical (DFT) measure of spin-phonon coupling: $\omega = \omega_0 + \lambda S_i S_j$ [36], $\lambda$ being negative. Using $<S_i S_j>$ for SNN spins obtained from Monte Carlo simulations, which is negative and sharply increases in magnitude below 100K [32], we confirm that the hardening of S1 mode below $T_s$ arises from the spin-phonon coupling and emergence of short-range stripe AFM2 order. As the ordering is short-range, we do not predict the splitting of S1 mode, as otherwise expected from a long-range AFM2 ordering. The coupling of $E_g$ and $B_{1g}$ modes with spin degree of freedom originates from the mode induced changes in Fe-Se-Fe bond-angle and consequent variation of the superexchange interactions. We note that a strong spin-phonon coupling has been inferred in iron pnictides ($CeFeAs_{1-x}P_xO$) through phenomenological analysis of the dependence of the measured magnetic moment of Fe on Fe-As layer separation and $T_c$ [15]. Our calculations show that the higher energy $E_g$ mode (S4 mode at 315 cm$^{-1}$ involving Fe displacements) also couples with spins with a comparable strength to the S1 mode. However, the perturbative analysis shows that its correction to frequency of S4 mode is much weaker (since the mode is of higher frequency, and correction depends inversely on the frequency of the mode).

## 4. CONCLUSION

To summarize, we have presented Raman measurements of FeSe$_{0.82}$, showing all the four Raman active modes and electronic Raman scattering involving 3$d$-orbitals of Fe. A picture consistent with our measured Raman spectra and DFT calculations is as follows: as the temperature is lowered, there is a transition to a phase with short range AFM2 stripe order at $T_s \sim 100$K, (as reflected in the anomaly observed in magnetic susceptibility [7]), long-range order being suppressed due to frustrated magnetic interactions. This change in spin ordering is accompanied by a weak orthorhombic distortion as a secondary order parameter. By symmetry, the coupling between this AFM2 order and $E_g$ (S1 and S4) modes is nonzero, which leads to anomalous change in frequency of the lower of the two $E_g$ modes (S1). Thus, a strong spin-phonon coupling (estimated from the phonon frequencies of AFM1 and AFM2 stripe phases) is responsible for the observed hardening of S1 mode in the *normal* state of FeSe. This may be relevant to the recent proposal of strong electron-phonon coupling through the spin-channel being discussed as a mechanism of superconductivity [15]. In addition, the high frequency modes at 1350 cm$^{-1}$ and 1600 cm$^{-1}$ are attributed to electronic Raman scattering involving $x^2$-$y^2$ to ($xz$, $yz$) $d$-orbitals of Fe. We conclude that spin polarized $xz$ and $yz$ orbitals of Fe, $E_g$ modes with in-plane displacement of Se atoms, orthorhombic strain and their couplings are relevant to the observed temperature dependent Raman spectra presented here.

**ACKNOWLEDGMENTS**

AKS and AKG acknowledge the DST, India, for financial support. PK, AK and JP acknowledge CSIR, India, for research fellowship. UVW acknowledge DAE Outstanding Researcher Fellowship for partial financial support.

TABLE I: Experimentally observed, calculated phonons frequencies and fitted parameters in cm$^{-1}$.

| Assignment | Experi--mental $\omega$ at 3K | Fitted parameters $\omega(0)$ | c | calculated Frequency x=0 | x=0.125 |
|---|---|---|---|---|---|
| S1 $E_g$ (Se) | 106 | $^a$101.5 ± 0.3 | -1.1±0.1 | 147 | 135 |
| S2 $A_{1g}$ (Se) | 160 | 161.8 ± 0.3 | -1.7±0.2 | 226 | 222 |
| S3 $B_{1g}$ (Fe) | 224 | 225.6 ± 0.4 | -3.2±0.4 | 251 | 239 |
| S4 $E_g$ (Fe) | 234 | 236.3 ± 0.1 | -2.1±0.1 | 315 | 304 |
| S5 | 254 | | | | |
| S6 CF (Fe) | 1350 | | | | |
| S7 CF (Fe) | 1600 | | | | |

$^a$Fitted Parameters to data from 100 to 300K.

**Figure caption:**

FIG.1. Powder x-ray diffraction pattern of $FeSe_{0.82}$ sintered at 720 ºC. Secondary hexagonal - FeSe (NiAs type) phase is marked by asterisk (*). Inset (a) Temperature dependence of resistivity (ρ) up to 200K (b) Temperature dependence of resistivity showing the criterion used to determine $T_C$ (c) Magnetic susceptibility as a function of temperature.

FIG. 2. (Color online): Unpolarized-Raman spectrum at 3K. Thick solid line (red) shows the total fit and thin solid lines (blue) show the individual fit. Inset (a) shows the S1 mode at two temperatures. Inset (b) shows the eigen-vectors of the calculated modes.

FIG. 3. (Color online): Temperature dependence of the modes S1, S2, S3 and S4. Solid lines are fitted lines as described in the text.

FIG. 4. (Color online): Crystal-field split energy level diagram for the Fe d-orbitals [see ref. 17 for energy values and ref. 18 for the electron distribution]. (a) Degeneracy of *xz* and *yz* orbitals is not lifted. (b) Degeneracy of *xz* and *yz* orbitals is lifted (Energy values are not to scale).

**Figure 1:**

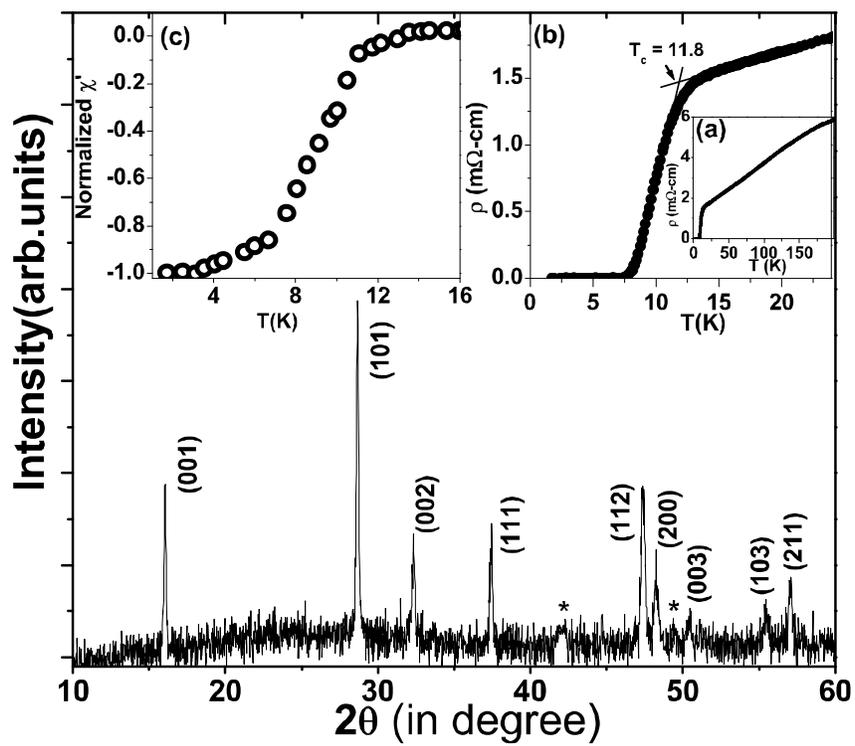

**Figure 2:**

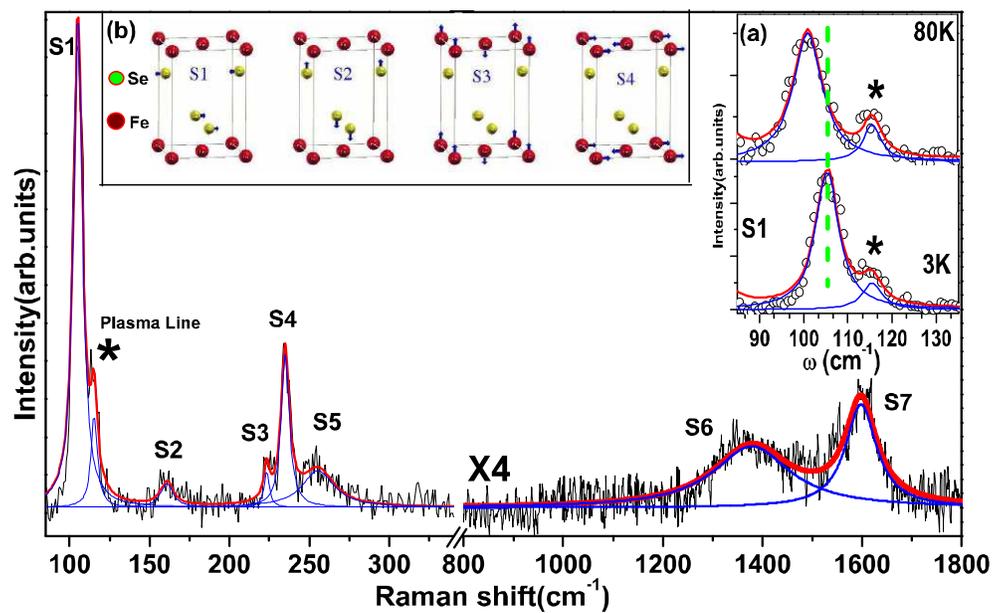

**Figure 3:**

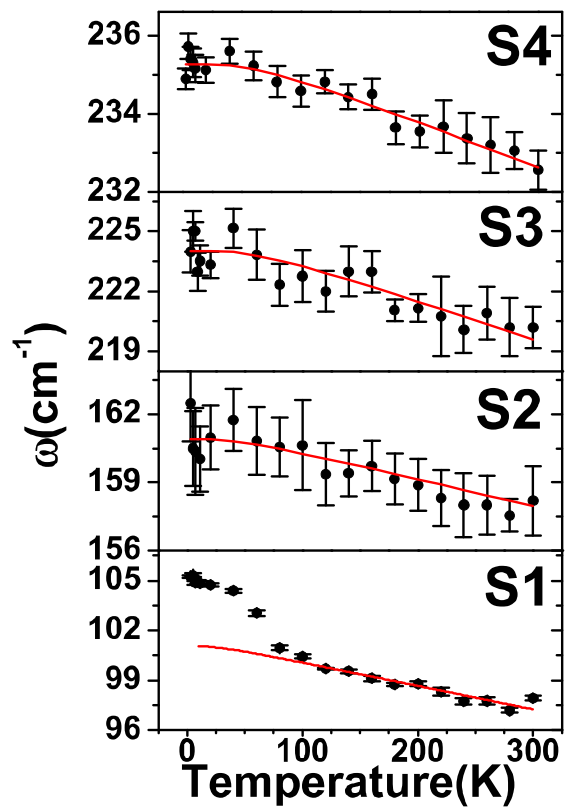

**Figure 4:**

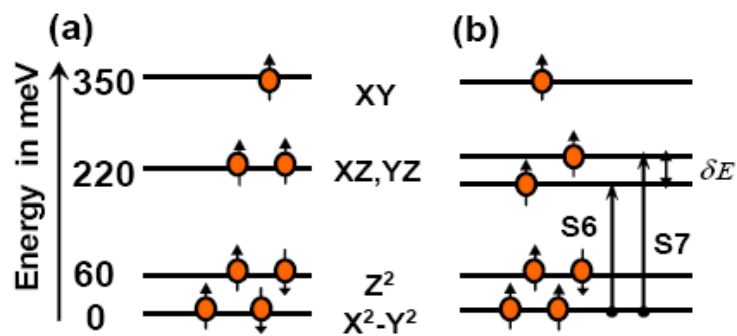